\documentclass[aps,prd,twocolumn,superscriptaddress]{revtex4-1}

\usepackage{graphicx}
\usepackage{color}
\newcommand{\w}{{\rm w}}
\begin{document}

\title{Radiative Thermal Noise for Transmissive Optics in Gravitational-Wave Detectors}


\author{Sheila Dwyer}
\email[]{dwyer\_s@ligo-wa.caltech.edu}
\affiliation{LIGO Hanford Observatory, PO Box 159, Richland, WA 99352, USA}
\author{Stefan W. Ballmer}
\email[]{sballmer@syr.edu}
\affiliation{Department of Physics, Syracuse University, NY 13244, USA}



\date{\today}

\begin{abstract}
Radiative losses have traditionally been neglected in the calculation of thermal noise of transmissive optical elements because for the most commonly used geometries they are small compared to losses due to thermal conduction.
We explore the use of such transmissive optical elements in extremely noise-sensitive environments such as the arm cavities of future gravitational-wave interferometers. This drives us to a geometry regime where radiative losses are no longer negligible. In this paper we derive the thermo-refractive noise associated with such radiative losses
and compare it to other known sources of thermal noise.
\end{abstract}


\maketitle

\section{Introduction}
The Gravitational wave interferometers currently under construction, Advanced LIGO \cite{Harry2010}, 
Advanced Virgo \cite{2013ASPC..467..151D} and Kagra \cite{Somiya:2011np} are expected to be limited by thermal noise in their most sensitive frequency band. This has driven a theoretical \cite{Braginsky1991, Braginsky2000, Levin1998, Levin2008, Levin2009, Evans2008} and experimental \cite{Harry:2001iw, Harry:06, Harry2010, Cole2013} interest in understanding and improving the fundamental thermal noise of optical elements in reflection and transmission. 

Thermal noise is a displacement noise. Its contribution to the strain sensitivity gets diluted with the arm length $L$. One of the most promising ways to overcome the sensitivity limitation due to thermal noise in future gravitational wave interferometers is to significantly increase the interferometer arm length from the current $4km$ (LIGO), or $3km$ (Virgo, Kagra). Inclusion of transmissive optical elements, such as lenses or wedges, in the interferometer arms, could make the design of such long interferometers more practical, at the cost of potentially introducing transmissive thermal noise.
We are therefore interested in the transmissive thermo-refractive noise in the limit of relatively large beams and thin optical elements.
Radiative losses as source for temperature fluctuations in thin films and semiconductors have previously been explored in van Vliet et. al. \cite{:/content/aip/journal/jap/51/6/10.1063/1.328104}, as well as van Vliet and Mehta \cite{PSSB:PSSB2221060102}. Braginsky et. al. \cite{coatingbook2012} 
and Heinert et. al. \cite{Heinert:11}
point out that such losses also lead to Stefan-Boltzmann radiation noise. 
For the most commonly used geometries radiative losses are however small compared to losses due to thermal conduction, and the literature then ignores radiative noise.
In the regime of large beam radii and thin optical elements this assumption is no longer true.  In this paper we derive the radiative thermo-refractive noise associated with such radiative losses, also referred to as Stefan-Boltzmann radiation noise.

The outline of this paper is as follows: Section \ref{TTRNintro} recapitulates the previously published transmissive thermo-refractive noise due to dissipation from thermal conduction. Section \ref{RTTRNcalc} gives the derivation of radiative transmissive thermo-refractive noise. This noise calculation is done using the Fluctuation-Dissipation theorem \cite{Callen1951}, which highlights the link between thermal noise and energy dissipation at a fundamental level.
Section \ref{compare} compares the result to other known noise sources, and we conclude in section \ref{conclusion}.

\section{Transmissive Thermo-Refractive Noise}
\label{TTRNintro}
The power spectral density of the transmissive thermo-refractive noise due to thermal diffusion was first calculated by Braginsky \cite{Braginsky2004} and is given by
\begin{equation}
\label{eq:oldTTRTN}
S_x(f) =  \beta_{eff}^2 a^2 S_{\delta T}(f) = \frac{16 k_B \kappa T^2 \beta_{eff}^2 a}{\pi C^2 \rho^2 \w^4 \omega^2} \,\, ,
\end{equation}
where $k_B$ is the Boltzmann constant, $\kappa$ is the thermal conductivity, $T$ is the temperature, $\beta_{eff}=dn/dT + (n-1)\alpha$ is the effective coupling accounting for the total phase shift due to a change in temperature, $\alpha$ is the thermal expansion coefficient, $a$ is the optic thickness, $C$ is the heat capacity per mass, $\rho$ is the density, $\w$ is the beam radius of the Gaussian read-out beam and $\omega=2 \pi f$ is the frequency.  Here $S_{\delta T}(f)$ is the temperature spectrum in the mode relevant for the optical readout. 
The derivation of equation \ref{eq:oldTTRTN} neglects radiative losses at the mirror surface. We will shown in this paper that those losses lead to additional noise given by
\begin{equation}
\label{eq:newTTRTNtop}
S_x(f) = \frac{16 k_B \epsilon \sigma T^5 \beta_{eff}^2}{\pi C^2 \rho^2 \w^2 \omega^2} \cdot N  \,\, .
\end{equation}
Here $\sigma$ is the Stefan–Boltzmann constant, $\epsilon$ is the emissivity and $N$ is the number of emitting surfaces, i.e. $N=2$ for a single transmissive element.

\section{Derivation}
\label{RTTRNcalc}

The transmissive thermo-refractive noise can be elegantly calculated  using the approach described by Levin \cite{Levin1998, Levin2008}: i) determine the read-out function for local thermal fluctuations, $q(\vec{x})$. ii) Apply a harmonic entropy (heating) profile of shape $q(\vec{x})$, frequency $f$ and arbitrary amplitude $F_0$. iii) Calculate the cycle-averaged dissipative losses $W_{\rm diss} = \frac{1}{\Delta t} \oint T \dot{S} dt$ for the geometry under consideration. The power spectral density of the associated thermal noise is then given by
\begin{equation}
\label{eq:MasterNoise}
S_{\delta T}(f) = \frac{8 k_B T}{\omega^2} \frac{W_{\rm diss}}{F_0^2}
\end{equation}

\noindent We can now apply this to the transmissive thermo-refractive noise:
\newline \noindent i) The measured path length fluctuation $\Delta x$ is given by
\begin{equation}
\Delta x = \left< \beta a  \delta T \right> = \beta_{eff} a \int_0^a dz \int dr^2  \delta T(z,\vec{r}) q(z,\vec{r})
\end{equation}
The readout function $q$ describes the weighting of fluctuations at different locations in the optic due to the average phase measurement. It is normalized to
\begin{equation}
\int_0^a dz \int dr^2  q(z,\vec{r}) = 1.
\end{equation}
If a Gaussian beam (field amplitude $\Psi \propto \exp{-\frac{r^2}{\w^2}}$) is used  for the read-out, we find for the read-out function q
\begin{equation}
q(z,\vec{r}) = \frac{1}{a} \frac{2}{\pi \w^2} e^{-2\frac{r^2}{\w^2}}
\end{equation}
We should point out here that we neglect any effects due to the standing-wave intensity pattern in a Fabry-Perot cavity. Benthem and Levin \cite{Levin2009} point out that this pattern leads to an increase of thermal noise once the diffusion length becomes comparable to the laser wavelength, which for fused silica fortunately happens above the gravitational wave observation band.

\noindent ii) We heat the test mass with 
\begin{equation}
\frac{dQ}{dV} = T ds = T F_0 \cos{(\omega t)} q(z,\vec{r})
\end{equation}
where $s$ is the entropy per unit volume. We need to solve the diffusion equation
\begin{equation}
C \rho \delta \dot{T} - \kappa \Delta \delta T = \frac{d}{dt} \frac{dQ}{dV} \,\, .
\end{equation}
For the application in mind the diffusion length $l_{th}=\sqrt{\kappa/(\omega C \rho)}$ is much smaller than the beam radius $\w$, $l_{th}<<\w$. For example the diffusion length for fused silica at the frequency of interest for gravitational wave interferometers (100Hz) is about $37 \mu m$, compared to a beam radius of several centimeters.
Since the beam radius determines the sharpest gradient in the heating profile, we can neglect the 2nd term and get for the temperature fluctuations
\begin{equation}
\label{eq:dT}
\delta {T} = \frac{1}{C \rho } \frac{dQ}{dV} =  \frac{1}{C \rho } T F_0 \cos{(\omega t)} q(z,\vec{r})
\end{equation}
iii) To calculate the dissipative loss $W_{\rm diss}$ we note that the rate of change of the entropy density is given by \cite{LandauLifshitz}
\begin{equation}
\dot{s}=-\frac{\vec{\nabla} \cdot \vec{j}}{T}
\end{equation}
where $\vec{j}$ is the heat flow per unit area. For the total increase of entropy we get
\begin{equation}
\dot{S}=-\int dV \frac{\vec{\nabla} \cdot \vec{j}}{T} = -\oint d\vec{A} \cdot \frac{\vec{j}}{T} + \int dV \vec{j} \cdot \vec{\nabla} (\frac{1}{T})
\end{equation}
The second term describes the thermal dissipation in the bulk of the material, which gives rise to the previously known thermo-refractive noise, while the surface term gives rise to the radiative thermo-refractive noise.  Using the fact that $\vec{j}=-\kappa \nabla T $, the 2nd term leads to the well-know dissipation formula
\begin{equation}
T\dot{S}_{\rm bulk}= - \int dV  \frac{1}{T}    \vec{j} \cdot \vec{\nabla} T = \int dV  \frac{\kappa}{T}    (\vec{\nabla} T)^2
\end{equation}
Applying it to the temperature fluctuations from equation \ref{eq:dT} yields the loss
\begin{equation}
W_{\rm diss} = \frac{1}{\Delta t} \oint T\dot{S}_{\rm bulk} dt = \frac{2 \kappa T F_0^2}{C^2 \rho^2} \frac{1}{a}  \frac{1}{\pi \w^4} 
\end{equation}
which, together with equation \ref{eq:MasterNoise}, gives the known thermo-refractive noise in equation \ref{eq:oldTTRTN}.

To calculate the contribution of the surface term we assume that the optic is in radiative thermal equilibrium with its surroundings, such that the average radiative heat flow is zero. We can again look at small temperature variations $\delta T$ around the average temperature $T$. For the surface term, we get
\begin{equation}
T \dot{S}_{\rm surf}=+ \frac{1}{T} \oint d\vec{A} \cdot  \vec{j} \delta T
\end{equation}
We can now use the linearized Stefan-Boltzmann law $\vec{j}=4 \epsilon \sigma T^3 \delta T$, with $\sigma$ the Stefan–Boltzmann constant and $\epsilon$ the emmissivity, to calculate the surface heat flow and average over one cycle of the fluctuation:
\begin{equation}
W_{\rm diss} = \frac{1}{\Delta t} \oint T\dot{S}_{\rm surf} dt = \frac{2 \epsilon \sigma T^4 F_0^2}{C^2 \rho^2} \frac{1}{a^2}  \frac{1}{\pi \w^2}  \cdot N
\end{equation}
where $N$ is the number of optical surfaces and $\Delta t$ is the period of the fluctuation.
This results in the radiative transmissive thermo-refractive noise of 
\begin{equation}
\label{eq:newTTRTN}
S_x(f) = \frac{16 k_B \epsilon \sigma T^5 \beta_{eff}^2}{\pi C^2 \rho^2 \w^2 \omega^2} \cdot N 
\end{equation}
To compare this to the conductive transmissive thermo-refractive noise (equation \ref{eq:oldTTRTN}) we calculate the total optic thickness for which the noises are equal
\begin{equation}
\label{eq:newTTRTN}
a_{\rm eq}= \frac{\epsilon \sigma T^3 \w^2}{\kappa} \cdot N = \frac{w^2}{l_{ch}} \cdot N
\end{equation}
For fused silica, the most promising material for transmissive elements due to its extraordinary low absorption, the characteristic length scale $l_{ch} = \kappa/(\epsilon \sigma T^3)$ is about $1~{\rm meter}$ at room temperature. 
As example, for an optical beam with radius $\w=7cm$ and two surfaces ($N=2$), the radiative and conductive transmissive thermo-refractive noise are equal for an optic thickness of $a_{\rm eq}=1cm$. In other words, if we imagine placing lenses  with a thickness of equal or less than about one centimeter in an interferometer arm, the radiative term would dominate the thermo-refractive thermal noise.

\section{Comparison to anti-reflective coating thermal noise}
\label{compare}
We  want to compare the radiative thermal noise to the coating thermal noise from an anti-reflective (AR) coating. First, for thermo-refractive noise we can refer to the calculation in \cite{Evans2008} and find
\begin{equation}
S_x(f) = \frac{2 \sqrt{2} k_B T^2}{\pi \w^2 \sqrt{\kappa C \rho \omega}}  (\beta_{eff}^{coat} d)^2 \cdot N 
\end{equation}
where $d$ is the coating thickness.  The effective coupling $\beta_{eff}^{coat}$ is given by the details of the coating. 
For a simple 1-layer AR coating we have
\begin{equation}
\beta_{eff}^{coat}= t_p \beta + (t_p n_c - 1) \alpha
\end{equation}
where $n_c$ is the coating index of refraction, and $t_p=\frac{d}{d\phi} (-i) \log t(\phi)$, where $t(\phi)$ is the coating transmission coefficient as function of the one-way propagation phase $\phi$ in the coating. For an ideal 1-layer AR coating with $n_c=\sqrt{n}$, $n$ the substrate index of refraction, $t_p$ is given by
\begin{equation}
t_p=\frac{1+n}{2\sqrt{n}}
\end{equation}
Note that for AR coatings this phase-enhancement in the coating does not need to be extreme, and $t_p$ is of order unity.
We then find that transmissive coating thermo-refractive noise is equal to radiative thermo-refractive noise for a coating thickness of
\begin{equation}
d_{\rm eq}=\left( \frac{4 \sqrt{2} \epsilon \sigma T^3}{\kappa} \right)^{1/2} \left(\sqrt{\frac{\kappa}{C \rho \omega}}\right)^{3/2} \frac{\beta_{eff}}{\beta_{eff}^{coat}}
\end{equation}
Again using the parameters for fused silica at room temperature, and a reference frequency of $100~{\rm Hz}$ we find $d_{\rm eq}\approx 0.5 \mu m \frac{\beta_{eff}}{\beta_{eff}^{coat}}$. That is the transmissive radiative and coating thermo-refracting noises are about equal in the band of interest of gravitational-wave detectors, with the radiative transmissive thermo-refractive noise dominating at lower frequencies.

Second, we would like to compare our result to the expected level of Brownian thermal noise, both due to internal and coating mechanical losses. To our knowledge there is no specific literature for transmissive Brownian noise. We therefore start with the known result for reflective Brownian noise \cite{Nakagawa:2001di,Harry:2001iw}
\begin{equation}
\label{coatingBrown}
\begin{array}{cc}
S_x(f) =  & \frac{2 k_B T}{\sqrt{\pi^3} f} \frac{1-\eta^2}{\w Y} 
\\  & \times \left( \phi_{\rm substrate}
 + \frac{2}{\sqrt{\pi}} \frac{1- 2 \eta}{1-\eta} \frac{d_{\rm coating}}{\w} \phi_{\rm coating} \right)
\end{array}
\end{equation}
where $d_{\rm coating}$ is the coating thickness, $\phi_{\rm substrate}$ and $\phi_{\rm coating}$ are the effective loss angles for substrate
and coating, and $\eta$ is Poisson's ratio for the substrate. The coupling to transmissive noise is reduced because 
(i) the coupling to optical path length is only $(n-1)$, and (ii) an AR coating is significantly thinner than an HR coating, resulting in less mechanically lossy material being used.
We should note though that for an optic thickness comparable or smaller than the beam radius, $a <\w$ equation \ref{coatingBrown} will break down, and the spot size scaling will change from $1/\w$ to $a/\w^2$.
Using parameters for fused silica,  $Y=72.8~{\rm Gpa}$, $\phi_{\rm substrate}=4 \cdot 10^{-10}$ \cite{Penn:2005jt} we find that the substrate Brownian noise is roughly equal to transmissive radiative thermo-refractive noise in the band around 100 Hz, but has a different dependence on spot size $\w$ and frequency $f$.
The Ti-doped ${\rm Ta_2O_5}$ coatings used in Advanced LIGO have a loss angle of about
$\phi_{\rm coating}=4.4 \cdot 10^{-4}$ \cite{Harry:06}.
Even with the reduced coupling to transmitted light and smaller coating thickness coating Brownian noise would still dominate the total noise. 
However,  if the coating loss can be reduced by a factor of 10, as demonstrated in \cite{Cole2013},  this will no longer be true.

\section{Conclusion}
\label{conclusion}
We showed that the previously published expression for transmissive thermo-refractive noise, equation \ref{eq:oldTTRTN}, needs a correction in the limit of large spot sizes and thin optical elements. This noise term, given in equation \ref{eq:newTTRTNtop}, is due to radiative losses. We gave a short derivation using Levin's approach \cite{Levin2008}. 
Using fused silica transmissive optical elements in a gravitational wave detector arm as an example, we compared the result to other know sources of thermal noise.  We found that in such a geometry transmissive thermo-refractive noise is of the same order of magnitude as coating thermo-refractive noise and substrate Brownian thermal noise. Coating Brownian thermal noise will still dominate the total noise, although this will no longer be true for crystalline coatings previously demonstrated in \cite{Cole2013}. 


\begin{acknowledgments}
We would like to thank Kiwamu Izumi and Daniel Sigg for many fruitful discussions. This work was supported by the National Science Foundation grants PHY-0823459 and PHY-1068809. This document has been assigned the LIGO Laboratory document number  LIGO-P1400123.
\end{acknowledgments}

\bibliography{LongIFO}
\bibliographystyle{plain}
\end{document}